# Extended generic nature of surface entropy

M. H. Ghatee*, A. Maleki, and H. Ghaed-Sharaf

*Department of Chemistry, Shiraz University, Shiraz, 71454, Iran, e-mail: ghatee@sun01.susc.ac.ir*

The recent empirical correlation for the surface tension is further investigated for universality. $C_5$-$C_8$ liquid hydrocarbons, for which accurate surface tension data over the complete liquid range are available, are used. Functional reduced surface entropy is derived, which is in the general form of Trouton's rule. It is characterized by the critical exponent of surface tension and is valid with the unique value 0.9990 at any temperature from freezing to critical temperature of the liquid hydrocarbons. Relation to generic nature of surface entropy is pursued.

As a part of our continuing investigation, previous works have led us to an empirical correlation for surface tension in which a family of linear curves indexed by reduced temperature $T^*_{index}$ are produced.[1,2] Highly correlated corresponding states correlations have been found for liquid hydrocarbons, liquid metals, and molten alkali halides. Having profound fundamental theories such as group renormalization[3,4], phenomenological scaling of the co-existent density, and MacLeod's equation for surface tension, which have given a statistical mechanical basis, strengthen the idea that such empirical behavior may corresponds to a concrete bulk or surface physical properties.[5,6]

In a more general way, its form consists of the functional reduced surface tension $\sigma^* = (\sigma/\sigma_f)(T_f/T)^{\mu}$ and the reduced temperature $T^* = [(\tau/\tau_f)(T_f/T)]^{\mu}$ where, $\sigma$ is the surface tension, T is the absolute temperature, and the subscript f refers to quantities at the freezing point. $\tau = (1 - T/T_c)$, with $T_c$ being the critical temperature. $\mu$ is the critical exponent for surface tension, and the theoretical value of 1.26, which is valid for the universality class involving liquid hydrocarbons under study in this work, is applied.[5,7,8]

Plot of $\sigma^*$ versus $T^*$ for $C_5$ to $C_8$ liquid hydrocarbons are shown in Figure 1. These are the only liquids for which accurate experimental surface tension data[9] are available from $T_f$ to $T_c$. For other liquid hydrocarbons data are available only at the low temperature limits and not are considered here. A promising linear curve with linear correlation coefficient 0.9999 indicates a perfect correlation. If the critical exponent $\mu$ is neglected in the definition of $T^*$, a non-linear curve smoothly passes through $\sigma^* - T^*$ data points. In this case, the slope varies between one and zero, when T is changed from $T_f$ to $T_c$, and keeps the track of singular part of the surface free energy as $T_c$ is approached. Thus, application of $\mu$ grantees the linear behavior of the correlation, specifically as $T_c$ is approached. This leads to an interesting application for the $\sigma^* - T^*$ correlation where accurate extrapolation of surface tension data up to $T_c$ is desired for a liquid of the same universality class.

The investigation continues by noting that, regardless of physical significant, the slope of the relation $\sigma^* = a + bT^*$ is quite constant from $T_f$ to $T_c$ and demonstrates the principle of corresponding states. Then, using $\sigma^* = \sigma^*(\sigma, T; \sigma_f, T_f, T_c, \mu)$ and $T^* = T^*(T; T_f, T_c, \mu)$, we have obtained,

$$\frac{d\sigma^*}{dT^*} = \tau\left(\frac{\tau}{\tau_f}\right)^{-\mu} \frac{1}{\alpha\,\sigma_f}\left[\sigma - \frac{T}{\mu}\left(\frac{d\sigma}{dT}\right)\right] \qquad (1)$$

where $\alpha$ is a constant order of unity, though slightly varies with temperature. The exact expression for $\alpha$ is derived by the application of the power law for surface tension of the form $\sigma = \sigma_o \tau^{\mu} + \tau^{0.5} + \cdots$, fitted to surface tension data by Grigoryov et al.[7-9] The temperature gradient of surface tension $(d\sigma/dT)$ is determined from the same power law.

A plot of $d\sigma^*/dT^*$ versus $\tau$ over the whole liquid range is shown in Figure 2. The functional of Figure 1 remarkably demonstrates a law of corresponding states. Deviations are shown by $C_7$, by about 0.50% to 0.75%, which may be attributed to the fact that $\tau_f$ of $C_7$ does not succeed the trend succeeded by $C_5$, $C_6$, and $C_8$.

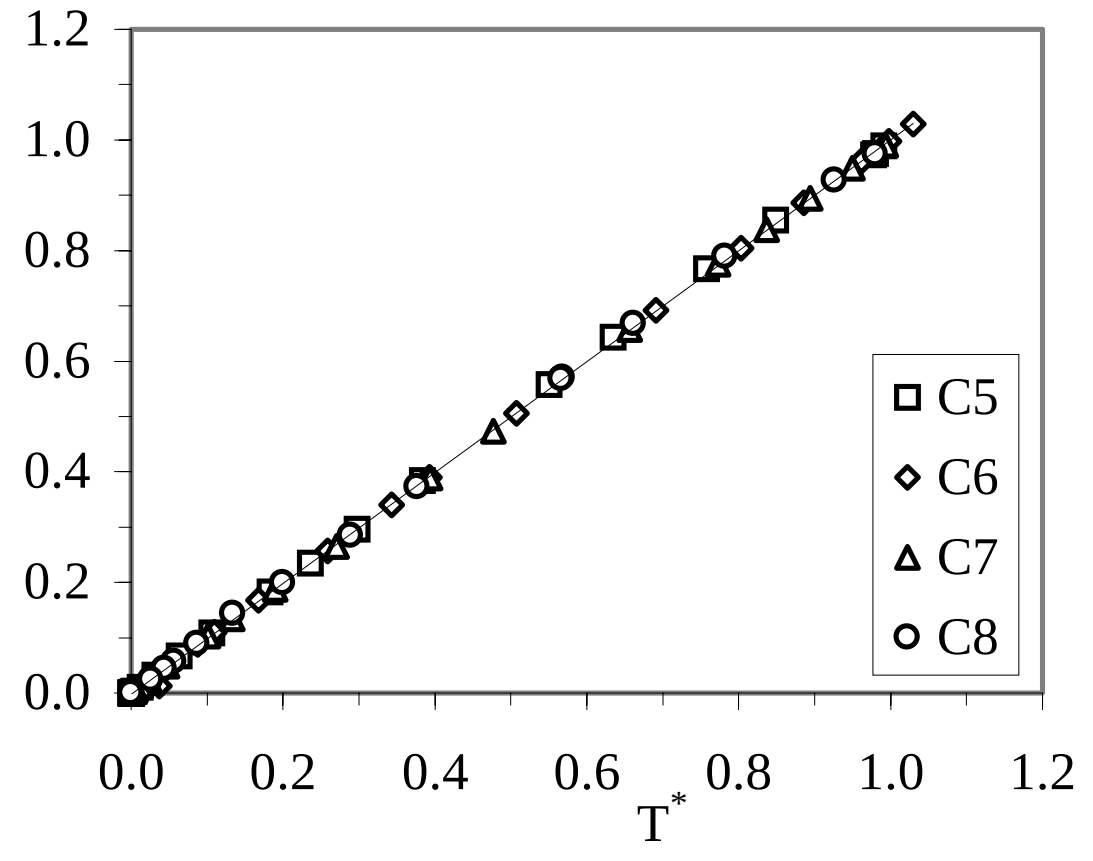

***Figure 1.*** $\sigma^*$ as a function of $T^*$ for $n-C_5H_{12}$, $n-C_6H_{14}$, $n-C_7H_{16}$, and $n-C_8H_{18}$ liquids.

The expression in the bracket of the equation (1) is the same as the expression for thermodynamic surface enthalpy, which is well approximated by surface energy $E^s$, except for the factor $1/\mu$ coupled with the surface entropy $(d\sigma/dT)$. Notice that the combination of first and second laws of thermodynamics leads to $E^s = [\sigma - T(d\sigma/dT)]$. Now we are in the position to express the physical significant of the slope of $\sigma^* - T^*$ correlation. From experimental observations and the results of this study, we propose

$$S^{s*} = E^{s*}_\mu / \tau^* \ (\cong \text{Universal value}) \qquad (2)$$

where,

$$\tau^* = \left(\tau^{\mu-1}/\tau_f^\mu\right) \qquad (3)$$

and,

$$E^{s*}_\mu = \frac{1}{\alpha}\left(E^s_\mu/\sigma_f\right), \qquad E^s_\mu = \left[\sigma - \frac{T}{\mu}\left(\frac{d\sigma}{dT}\right)\right] \qquad (4)$$

is the same value for all systems of the same universality class. $E^s_\mu$ is the surface energy defined by equation (4) and the subscript $\mu$ emphasis the additional $1/\mu$ factor with respect to $E^s$. The expression (2) for $S^{s*}$ is original. This expression is not as straightforward as the Trouton's rule for normal liquids, but pertains the same concept. It states that the quotient of reduced surface energy $E^{s*}_\mu$ and the temperature scaled from $T_f$ and $T_c$, $\tau^*$, is a constant over the whole liquid range of liquids considered here. Additionally, in a more general way, **1)** expression (2) is a constant independent of the temperature for liquids of the universality class, and **2)** it is valid over the whole liquid range including critical region up to $\tau \approx 10^{-5}$.

It is useful to note that the law of corresponding states is applied here if one express the surface tension at the freezing

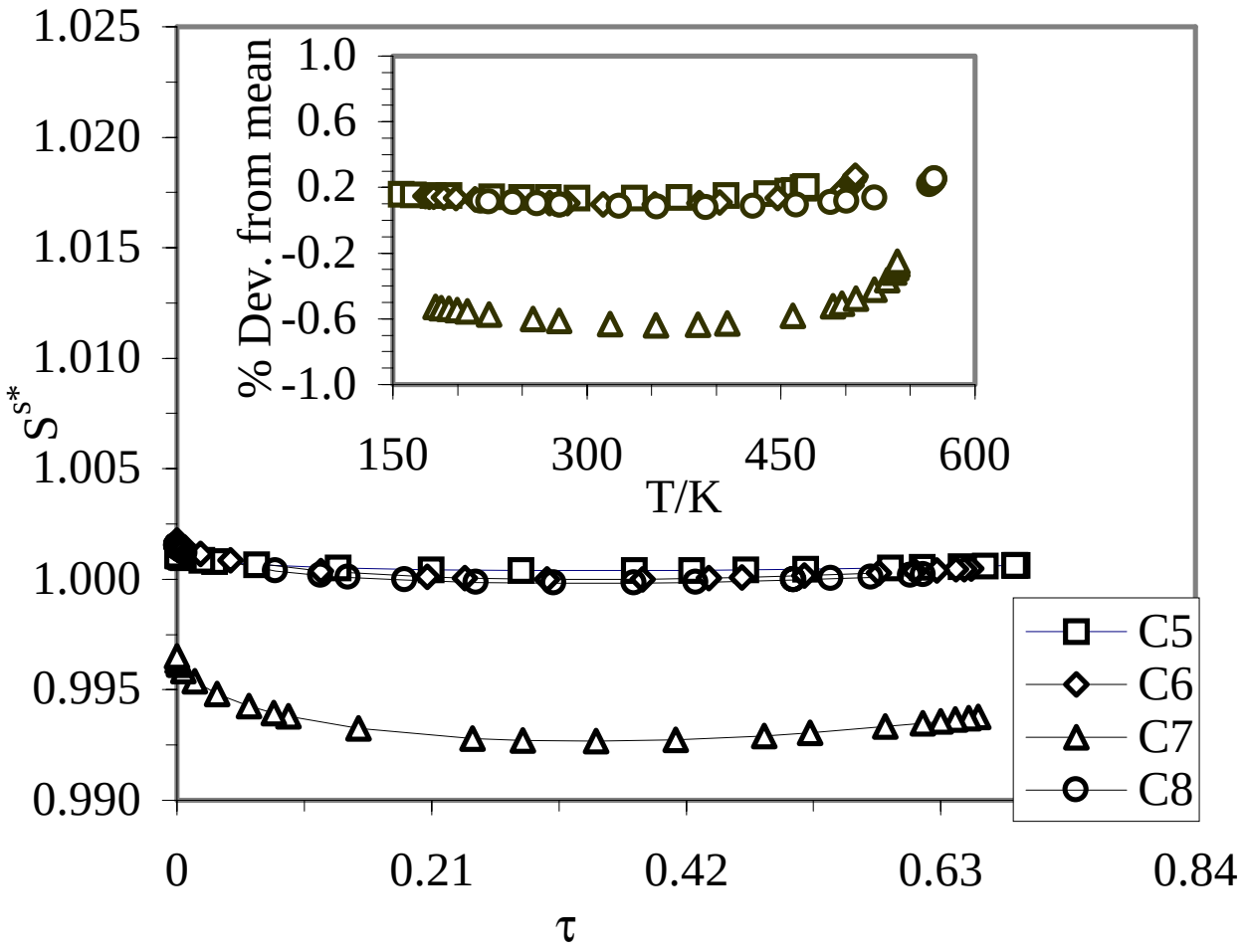

***Figure 2.*** $S^{s*}$ versus $\tau$. for the same systems as in Figure 1. Insert shows the percent deviations from mean values of $S^{s*}$.

point in the form

$$\sigma_f = a_\circ^2 (2B_\circ \rho_c)\, \tau_f^{(\beta+\varphi)} \qquad (5)$$

where $\rho_c$ is the critical density, $a_\circ^2$ and $B_\circ$ are the amplitude of capillary constant and the amplitude of $\Delta\rho$ of the liquid-vapor system that satisfies the corresponding power laws, respectively. The critical exponent of the later one $\beta = 0.325$, and the former one $\varphi = 0.935$. Thus, $S^{s*}$ would be independent of freezing physical properties if one substitute (5) in (4). Then, the conventional law of corresponding states for $S^{s*}$ follows by having the surface energy reduced by appropriate combination of critical parameter, e.g. $\rho_c$ and $T_c$:

$$S^{s*} = \frac{1}{2a_\circ^2 \alpha \rho_c B_\circ}\left(\frac{E^s_\mu}{\tau^{\mu-1}}\right) \qquad (6)$$

Liquid hydrocarbons $C_5$ to $C_8$ accurately follow Eq (2) over the whole liquid range, with the average value $S^{s*} = 0.9990$. The deviation plot inserted in Figure 2 indicates $-0.64\% \rangle \Delta S^{s*} \rangle 0.26\%$, with the maximum deviation due to $C_7$.

The surface tension is representing the surface free energy per unit area, and thus, it is contributed from both surface energy, and surface entropy. Considering the surface energy of numerous liquid hydrocarbons, molten salts, and molten metals, it has been shown that surface energy changes by about three order of magnitude, while surface entropy is within certain limits and essentially remains constant within experimental error. It is concluded that, surface energy is a liquid-specific property and surface entropy a liquid-generic properties.[10] Perhaps, it implies that the surface energy is defined at the distance far away to the liquid side of the Gibbs dividing surface, whereas the surface entropy is dominated by the middle part.[11] On the liquid side, the mean free path is comparable to the range of intermolecular interaction and thus, the internal energy i.e., the surface energy is influenced by the nature of liquid. Certainly, as the critical point is approached, the increase in the fluctuation of density will cause a large fluctuation in the position of the dividing surface. Thus,

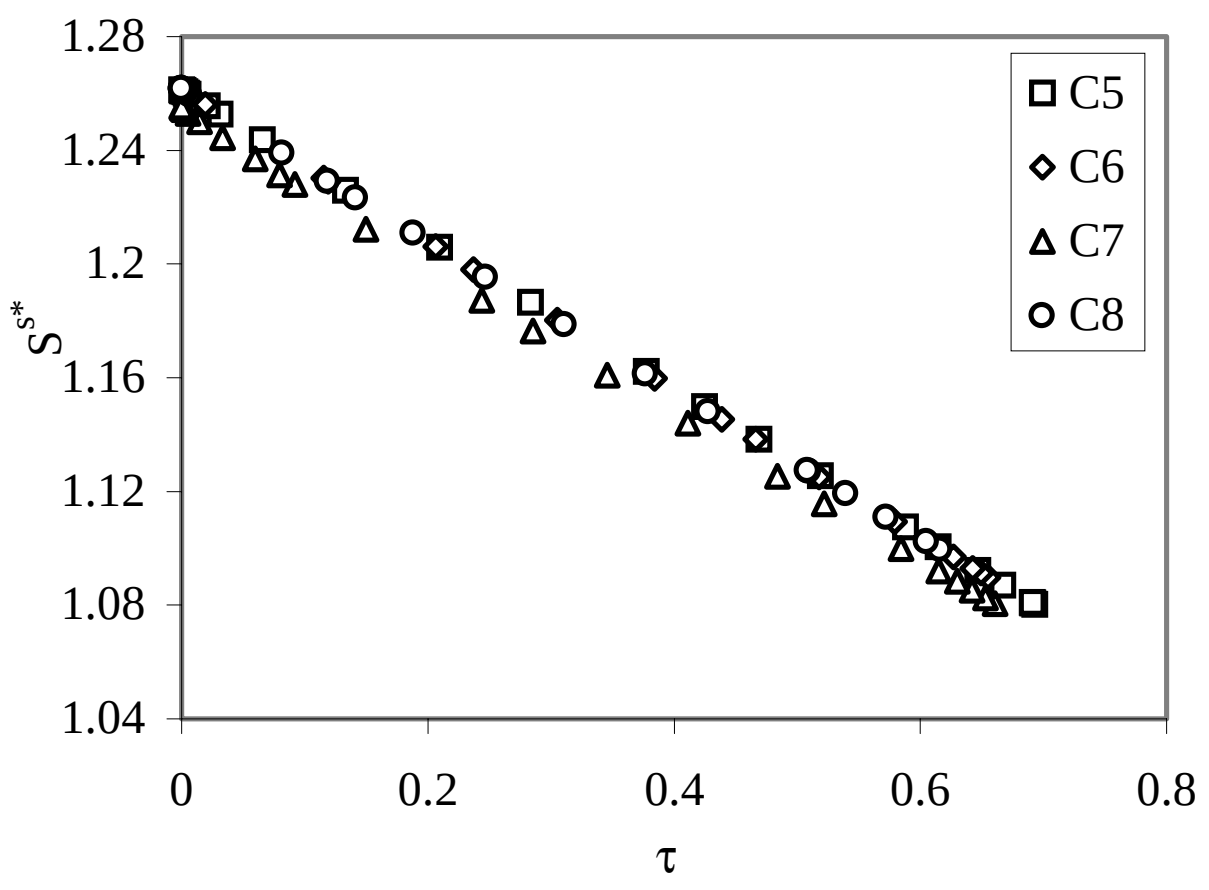

***Figure 3***. The same plots as in Figure 2 except $E^s_\mu$ replaced by $E^s$ in expression (2).

the singularity in the surface free energy in this region might be attributed to the contribution of surface energy as well as surface entropy. While the surface entropy is subjected to the behavior of the middle range of the interfacial region rigorously, close enough to the critical temperature, the singularity in the surface free energy has to be accounted by the critical exponent $\mu$ to be certain the law of corresponding states holds. This explains the reason for the definition of $\sigma^*$ and $T^*$ in this work to mimic the concept embedded by the Trouton's rule

In summary, we have developed analytical expression (2) and (6) by considering the theoretical result for density and surface tension of coexistence liquid-vapor system. The surface energy could mimic the Trouton's rule in a corresponding states manner, and applies quite well to a number or of liquid hydrocarbons for which surface tension data are available from $T_f$ up to $\tau \approx 10^{-5}$. The rule includes the characteristic surface energy $E^s_\mu$.

Figure 3 shows the same plots as in Figure 2 except $E_i^s$ replaced by $E^s$. It can be seen that the uniqueness of $S^{s*}$, as persisted in Figure 2, is totally lost.

In conclusion, the application of expression (2) [equivalently expression (6)] and the consequent of Figure 3 rise the question how could the factor $1/\mu$ be essential in $E_i^s$ for substantiating the corresponding states correlation. In other word, could one promote a combination of first law and second law of thermodynamic, which can yield surface energy as $E_i^s$? The question influences the answer we are seeking for in that the critical exponent $\mu$ and surface entropy $(d\sigma/dT)$ are both generic properties of fluids of universality classes.

**Acknowledgement.** The authors are indebted to the research council of Shiraz University for supporting this project.